\documentstyle[prl,aps,preprint,epsf]{revtex}
\tightenlines 
\begin{document}
\draft 


\title{Activated
escape of periodically driven systems}

\author{M.I. Dykman$^1$\cite{byline}, B. Golding$^1$, L.I.~McCann$^{1,2}$,
V.~N.~Smelyanskiy$^{1,3}$\\
{\it and}\\ 
D.~G.~Luchinsky$^{4}$, R.~Mannella$^{4,5}$, and
P.~V.~E.~McClintock$^{4}$}

\address{
$^1$Department of Physics and Astronomy, Michigan State University,
East Lansing, MI 48824\\
$^2$Department of Physics, University of Wisconsin-River Falls, River Falls, WI 54022\\
$^3$NASA Ames Research Center, MS 269-2, Moffett Field, 
CA 94035-1000\\
$^4$Department of Physics, Lancaster University, Lancaster
LA1 4YB, UK\\
$^5$Dipartimento di Fisica, Universit\`{a} di Pisa and INFM UdR Pisa,
Via F.~Buonarroti~2, 56100 Pisa, Italy}

\maketitle 

\begin{abstract}
We discuss activated escape from a metastable state of a system driven
by a time-periodic force. We show that the escape probabilities can be
changed very strongly even by a comparatively weak force. In a broad
parameter range, the activation energy of escape depends linearly on
the force amplitude. This dependence is described by the logarithmic
susceptibility, which is analyzed theoretically and through analog and
digital simulations. A closed-form explicit expression for the escape
rate of an overdamped Brownian particle is presented and shown to be
in quantitative agreement with the simulations.  We also describe
experiments on a Brownian particle optically trapped in a double-well
potential.  A suitable periodic modulation of the optical intensity
breaks the spatio-temporal symmetry of an otherwise spatially
symmetric system. This has allowed us to localize a particle in one of
the symmetric wells.
\end{abstract}


\noindent
{\bf Fluctuation-induced escape from a metastable state is at the root
of many physical phenomena, from diffusion in crystals to protein
folding, and is closely related to nucleation in phase transitions and
activated chemical reactions.  In all these phenomena it would be
advantageous to control the escape probability by applying an external
force. The problem of escape of driven systems has therefore attracted
much attention in diverse contexts, a recent application being stochastic
resonance \cite{SR-reviews}. We show that this problem can be solved
in a very general form for a broad range of driving field frequencies,
which goes far beyond the adiabatic limit. The analytic theory is
compared with the results of analog and digital simulations.  We then
discuss experiments on controlling escape in modulated optical
traps. An important application of the results is the possibility of
selective control of particles diffusion in a periodic potential,
including both the rate and direction of the diffusion.}

\section{Introduction}

The question of how a system responds to an external field is one
of the fundamental problems of physics. A strong nonlinear response is
usually associated with a sharply resonant excitation of the
system. However, the effect of external driving may also be extremely
large for an important and wide class of phenomena related to large
fluctuations, including escape from a metastable state and
nucleation in phase transitions.

The mechanism responsible is readily understood for adiabatically slow
driving, where the driving frequency is small compared to the
relaxation rate  in the absence of fluctuations and the system remains in
quasi-equilibrium. For systems in thermal equilibrium, the fluctuation
probabilities are given by the activation law, $W\propto
\exp(-R/k_BT)$. For large infrequent fluctuations, which are discussed
in the present paper, the probabilities $W$ are much less than all
frequencies and relaxation rates. We
will be specifically interested in activated escape, in which case $R$
is the activation energy of escape. The driving force modulates the
value of $R$ quasi-statically and, even where the modulation amplitude
$|\delta R|$ is small compared to $R$, it may still substantially
exceed $k_BT$, in which case $W$ will be changed very strongly. We
emphasize that the change of the activation energy is {\it linear} in
the field amplitude, for $|\delta R|\ll R$.

For higher field frequencies, where the driving becomes nonadiabatic,
the expected major effect of the field would be to ``heat up'' the
system by changing its effective temperature. Indeed, in the
weak-field limit, the escape rate $W$ is known, theoretically
\cite{Larkin} and experimentally \cite{Devoret}, to be incremented by
a term proportional to the field {\it intensity} $I$ rather than the
{\it amplitude} $A\propto I^{1/2}$ \cite{Larkin}. However, one may ask
what happens if the appropriately weighted field amplitude is not
small compared to the fluctuation intensity (temperature), and whether
an exponentially strong change of the escape rate will occur.

Theoretical analysis of nonadiabatically driven systems is
complicated, since one may no longer assume that the system is in
thermal equilibrium.  Whereas for equilibrium systems the exponent in
the escape rate can be found, at least in principle, as the height of
the free-energy barrier, for nonequilibrium systems there are no
universal relations from which it can be obtained \cite{Landauer};
the situation with the prefactor is even more complicated
\cite{MS_prefactor}. Much effort has been put into solving the
nonadiabatic response problem, in diverse contexts, and numerical
results have been obtained for specific models (see e.g.
\cite{Jung:93,Reichl:96}).

Recent theoretical results \cite{Dykman:97a,Dykman:97b} show that,
counter-intuitively, for high-frequency driving the change of $R$ is
proportional to the field amplitude, i.e., $\ln W$ is linear in $A$,
over a broad range of $A$. The proportionality coefficient was called
the logarithmic susceptibility (LS). Just like the conventional linear
susceptibility, the LS relates the response of the system in the
presence of external driving to its dynamics in thermal equilibrium in
the absence of the driving field. We emphasize that the amplitude is a
nonanalytic characteristic of the field, as it is obtained by taking
the square root of the period-averaged squared field. We are therefore
talking about a nonanalytic field dependence of the escape rate, and
we need to determine a mechanism that would lead to such a dependence.

In Sec.~II we provide a general formulation which allows one to find,
for a periodically driven system, the activation energy of escape
induced by Gaussian noise with an arbitrary power spectrum. In Sec.~III
we outline the theory and analyze the frequency dispersion of the
LS. We then discuss the results on the prefactor in the escape rate of
a driven system \cite{SDG} and analyze the full time-dependent as well
as the time-averaged escape rate, including both the exponent and
prefactor. In Sec.~IV we present the results of analog and digital
simulations of driven systems. These results provide a full
qualitative and quantitative confirmation of the theory, and also
reveal the underlying physics explicitly. In Sec.~V we describe the
experimental observations of the activated escape of particles in
modulated optical traps.  Sec.~VI contains conclusions and a discussion
of unsolved problems in activated escape, including the problem of
statistical reconstruction of the dynamical model of a fluctuating
system.

\section{General formulation of the escape problem}

The idea underlying the theory of the LS \cite{Dykman:97a,Dykman:97b}
is that, although the motion of the fluctuating system is random, in a
large rare fluctuation from a metastable state to a remote state, or
in a fluctuation resulting in escape, the system is most likely to
move along a particular trajectory known as the optimal path (see
\cite{Onsager,Freidlin:84,Dykman:84,Graham:89,Luciani,Mikhailov,Bray:89,Dykman90,M&S}
and references therein). The effect of the driving field accumulates
as the system moves along the corresponding optimal path, giving rise
to a linear-in-the-field correction to the activation energy of
escape.

A natural theoretical approach to the escape problem is based on the
path-integral technique. We will give a  formulation which
is based on this technique and allows one to find the logarithm of the
escape rate for a {\it periodically driven system}. We consider a
general case where fluctuations in the system are caused by a
stationary colored Gaussian noise $f(t)$ with a power spectrum
$\Phi(\omega)$ of arbitrary shape \cite{Dykman90,Dykman99}. The
Langevin equation of motion is of the form:
\begin{equation}
\label{Langevin}
\dot q = K(q;t) + f(t),\quad K(q;t+\tau_F) = K(q;t),
\end{equation}
\noindent
where $\tau_F$ is the period of the driving field. The noise is fully
characterized by its correlation function $\phi(t)= \langle f(t)f(0)
\rangle$ or by $\Phi(\omega)$, the Fourier transform of $\phi(t)$. The
characteristic noise intensity is $D=\max\Phi(\omega)/2$.

If the noise is weak then, over the noise correlation time $t_{\rm
corr}$ and the characteristic relaxation time in the absence of noise
$t_{\rm rel}$, the system will approach the metastable periodic state
$q_a(t)$ and will then perform small fluctuations about it
\cite{Gitterman}. To escape from the basin of attraction of this
state, the system should be subjected to a sufficiently large pulse of
the force $f(t)$. Various realizations of $f(t)$ [the pulse shapes]
can result in escape. Their probability densities are given by the
functional \cite{Feynman}
\begin{equation}
\label{prob_fnctnl}
{\cal P}[{f}(t)] = \exp\left[-{1\over 2D}\int dt\,dt'\,
f(t)\hat{\cal F}(t-t')f(t')\right],
\end{equation}
\noindent 
where $\hat {\cal F}(t)$ is a reciprocal of the noise correlation
function $\phi(t)$, $ \int dt_1\,\hat {\cal F}(t-t_1)\phi(t_1-t') =
D\delta(t-t')$. For white noise, $\hat{\cal F}(t) = \phi(t)/4D
=\delta(t)/2$.

We assume that the noise intensity $D$ contains a small constant,
which is the small parameter of the theory. This parameter guarantees
that the functional (\ref{prob_fnctnl}) is exponentially small for all
pulses $f(t)$ which can give rise to escape. In addition, its values
differ exponentially for different appropriate $f(t)$. Thus there
exists a realization $f(t)=f_{\rm opt}(t)$ which is exponentially more
probable than the others. This optimal realization provides the
maximum to ${\cal P}$ subject to the {\it constraint} that the system
(\ref{Langevin}) actually escapes.  The path $q_{\rm opt}(t)$ along
which the system moves when driven by the optimal force $f_{\rm
opt}(t)$ is the optimal fluctuational path, $q_{\rm opt}(t)$.

From (\ref{prob_fnctnl}), the paths $q_{\rm opt}$, $f_{\rm opt}$
provide the minimum to the functional
\begin{eqnarray}
{\cal R}[q(t),f(t)] = &&{1\over 2}\int\!\!\int_{-\infty}^{\infty} dt\,dt'\,
f(t)\hat {\cal F}(t-t')f(t')
+ \int_{-\infty}^{\infty} dt\, \lambda(t)
\left[\dot q - K(q;t)-f(t)\right].
\label{varfunct}
\end{eqnarray}
\noindent 
They can be obtained from the corresponding variational equations of
motion. The Lagrange multiplier $\lambda(t)$ relates
$f_{\rm opt}(t)$ and $q_{\rm opt}(t)$ to each other.

The boundary conditions for the escape problem follow from the fact
that the system starts from the periodic attractor
$q_a(t)=q_a(t+\tau_F)$ in the distant past (on the order of the
reciprocal escape rate), with $f=0$ asymptotically, and that, as the
force decays after having driven the system away from the attractor,
the system should not be brought back to the initially occupied basin
of attraction. The latter condition is only satisfied \cite{Dykman90}
if, for $t\to\infty$, the system is approaching the unstable periodic
state $q_b(t)=q_b(t+\tau_F)$ on the boundary of the basin of
attraction to $q_a(t)$,

\begin{eqnarray}
&&f(t) \rightarrow 0,\; \lambda(t)\to 0 
\quad {\rm for} \;
t\rightarrow \pm \infty;\nonumber
\\ 
&& 
 q(t) \rightarrow q_a(t) 
\; {\rm for}
\; t\rightarrow - \infty; \quad q(t) \to q_b(t)
\; {\rm for} \;
\; t\rightarrow \infty. 
\label{boundary}
\end{eqnarray}

The time-averaged escape rate has the form
\begin{equation}
\label{escape_rate}
\bar W =C \exp[-R/D], \; R = \min{\cal R}.
\end{equation}
The exponent $R$ can be obtained for an arbitrary noise spectrum and
an arbitrary periodic driving by solving the variational problem
(\ref{varfunct}), (\ref{boundary}) numerically. In particular, in the
case of white noise, where $\hat{\cal F}(t)=\delta (t)/2$, the
Lagrange multiplier and the force $f(t)$ can be easily eliminated from
the variational equations, $2\lambda(t)=f(t)=\dot q - K$, and the variational
functional ${\cal R}$ for the escape problem takes the form
(cf. \cite{Dykman:97b})

\begin{equation}
\label{white_noise_R}
{\cal R}[q(t)]={1\over 4}\int_{-\infty}^{\infty}dt\, [\dot q - K(q;t)]^2.
\end{equation}

The variational equations of motion for the problem (\ref{varfunct})
are usually nonintegrable. In the case of a white-noise driven system
this was pointed out by Graham and T\'el \cite{Graham:89}. Generically
there are several solutions which start from the attractor for $t\to
-\infty$ and arrive to a given state $q_f$ at a given time $t_f$. The
physically meaningful observable solution $q_{\rm opt}(t)$ provides the {\it
absolute minimum} to the functional ${\cal R}$ \cite{DMS}.

The prefactor $C$ in the escape rate (\ref{escape_rate}) and the
relation of $W$ to a directly observable quantity, the time-periodic
current from the basin of attraction, are discussed below. 

\section{The logarithmic susceptibility}

We now turn to the case where the driving force $F(t)$ is additive, 
\begin{equation}
\label{force}
K(q;t) = -U'(q) + F(t), \; F(t+\tau_F)=F(t),
\end{equation}
and only weakly perturbs the system dynamics; in particular, it does
not change the number of attractors or saddle states. Even in this
case the effect of $F(t)$ on the escape probability may be
exponentially strong \cite{Dykman:97a,Dykman:97b,SDG}, because it is
determined by the {\it ratio} of the field-induced increment $\delta
R$ of the escape activation energy to small noise intensity $D$. We
note that $U(q)$ can be thought of as a metastable potential in which the
system moves in the absence of periodic driving.

To first order in $F$, the correction $\delta R$ can be obtained from
the variational functional (\ref{varfunct}) by evaluating the term
$\propto F(t)$ along the zeroth-order path $q_{\rm
opt}^{(0)}(t),f_{\rm opt}^{(0)}(t)$, $\lambda^{(0)}(t)$. However,
special care has to be taken of the fact that the optimal escape path
is an {\it instanton} \cite{Dykman:97b}. In particular, the function
$\lambda^{(0)}(t)$ is other than zero within a time interval of width
$\sim t_{\rm corr},\, t_{\rm rel}$ and is exponentially small
otherwise. At the same time, the optimal fluctuation leading to escape
may occur at any time $t_c$, in the absence of periodic driving (one
can think of $t_c$ as the ``center'' of the instanton where
$|\lambda^{(0)}(t)|$ reaches its maximum).

The field $F(t)$ lifts the time degeneracy of escape paths. It {\em
synchronizes} optimal escape trajectories, one per period, so as to
minimize the activation energy of escape $R$. The field-induced change
of $R$ should be evaluated along such a trajectory, i.e.

\begin{eqnarray}
\label{escape_chi}
\delta R = &&\min_{t_c}\delta R(t_c),\quad \delta R(t_c)=
\int_{-\infty}^{\infty} dt\,
\chi(t-t_c)F(t)\nonumber\\
&&\equiv
\sum_n\tilde\chi(n\omega_F)F_n\exp(in\omega_Ft_c),\quad \chi(t)=
-\lambda^{(0)}(t),
\end{eqnarray}
where $\tilde\chi(\omega)=\int_{-\infty}^{\infty}dt\,
\chi(t)\exp(i\omega t)$, and $F_n$ is the $n$th Fourier component of
the field \cite{Dykman99}.  A complete derivation for a white-noise
driven system is discussed in Ref.~\cite{Dykman:97b}; for the general
case discussed here it will be given elsewhere. We note that
Eq.~(\ref{escape_chi}) has a particularly simple form for sinusoidal
driving, $F(t) = A\cos \omega_Ft$. In this case $\delta R = -|\tilde
\chi(\omega_F)|A$.

The change of the activation energy $\delta R$, and therefore the
logarithm of the escape rate $\bar W$, are linear in
the field $F(t)$. The coefficient $\tilde\chi(\omega)$
%
%
is the {\it logarithmic susceptibility} (LS)
\cite{Dykman:97a,Dykman:97b}. The function $\chi$ is a characteristic
of the system, as are, for example, the polarizability and other
standard linear susceptibilities. It can be calculated for a given
model or measured experimentally. 

Unlike the standard linear susceptibility which, by causality
arguments, is given by a Fourier integral over time from $0$ to
$\infty$, $\tilde\chi(\omega)$ is given by an integral from $-\infty$
to $\infty$. The analytic properties of $\tilde\chi(\omega)$ therefore
differ from those of the standard susceptibility and, in particular,
their high-frequency asymptotics are {\it qualitatively}
different. The standard susceptibility for a damped dynamical system
decays as a power law for large $\omega$ [e.g., as
$1/[U^{\prime\prime}(q_a) -i\omega]$, for the model (\ref{force})]. In
contrast, from (\ref{escape_chi}) the LS decreases {\it
exponentially}, $\tilde\chi(\omega)=M e^{-|\omega|\tau_p}$, where
$\tau_p = {\rm Im}\,t_p$, and $t_p$ is the pole or the branching point
of the function $\lambda^{(0)}(t)$. The asymptotic behavior of
$\tilde\chi(\omega)$ is different, of course, if $\tau_p=0$,
i.e. $\lambda^{(0)}(t)$ has a singularity for real time. This happens,
for example, if the potential $U(q)$ has singularities encountered by
the optimal path. Therefore it does not typically occur in dynamical
systems.

The LS takes a particularly simple form for a white-noise driven
system. From (\ref{white_noise_R}), (\ref{force}),
\begin{equation}
\chi(t)=-\lambda^{(0)}(t)=- \dot q^{(0)}_{\rm opt}(t),\quad \dot
q^{(0)}_{\rm opt} = U^{\prime}(q^{(0)}_{\rm opt}).
\label{white-explicit}
\end{equation}
In this case, the explicit form of $\tau_p$ and the prefactor $M$ in
$\tilde\chi(\omega)$ are determined solely by the singularities of
$U'(q)$. They were obtained in Ref.~\cite{Luch:jpa}.

\subsection{Complete nonadiabatic escape theory}
The notion of the LS makes it possible to find not only
the exponent, but also the prefactor in the escape rate, and thus to
obtain a complete nonadiabatic solution of the escape problem for
dynamically weak driving. Since the celebrated Kramers paper
\cite{Kramers}, the calculation of the prefactor has been one of the
central problems in escape rate theory. For a periodically driven
system, the escape rate $W(t)$ is periodic in time. It can be
introduced as a current $j$ away from the metastable state, which is
measured {\it well behind} the boundary $q_b$ of the attraction basin
[for the model (\ref{force}) with $F=0$, $q_b$ is the position of the local
maximum of the potential $U(q)$]. In the range $|U'(q)|\gg F$ the
current scales with $q$ as

\begin{eqnarray}
\label{current}
&&j(q,t) = W[t-t_d(q)],
\; dt_d/dq =- 1/U^{\prime}(q).
\end{eqnarray}

Eq.~(\ref{current}) provides a meaningful definition of both
instantaneous and time-averaged escape rates. For weak driving, the
values of the escape rate at different points $q$ sufficiently far
behind $q_b(t)$ differ only by a phase shift $t_d(q)$, which makes it
possible to make a sensible measurement of $W(t)=W(t+\tau_F)$.  For a
white-noise driven system, an explicit expression for the
time-dependent escape rate $W(t)$ and for $\bar W$ was obtained
\cite{SDG} by combining the results on the LS with the integral
representation of the time-dependent probability density near
$q_b$. In particular, it was shown that

\begin{eqnarray}
\label{average}
\bar W/ W_0 = (2\pi)^{-1}\int_0^{2\pi}d\phi\,\exp[-\delta R(\phi/\omega_F)/D],
\end{eqnarray}
where $W_0$ is Kramers' escape rate in the absence of modulation for
an overdamped system (the type of systems which we discuss in this
paper), and $\delta R(t_c)$ is given by Eq.~(\ref{escape_chi}).

\begin{figure}
\epsfxsize=3.5in                
\leavevmode\epsfbox{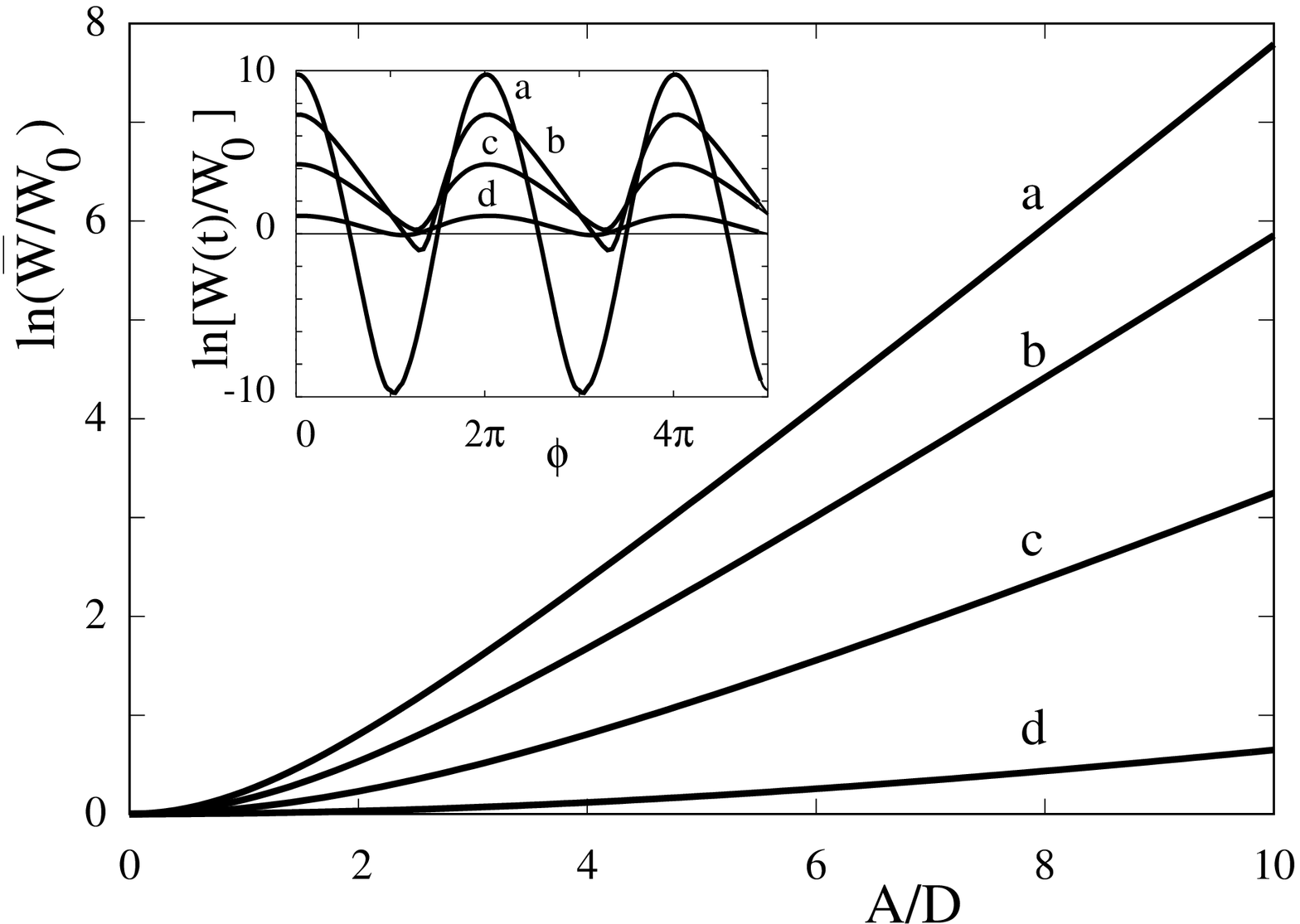}
\vspace{0.1in}
\caption{The logarithm of the average escape rate
(\protect\ref{average}) as a function of the scaled amplitude $A/D$ of
a sinusoidal field for the potential $U(q) = q^2/2-q^3/3$
\protect\cite{SDG}. The curves $a$ to $d$ refer to the dimensionless
frequency $\omega_F=0.1, 0.4, 0.7, 1.2$. Inset: time dependence of the
{\it logarithm} of the instantaneous escape rate for the same
frequencies and $A/D=10$ ($\phi = \omega_Ft$), illustrating loss of
synchronization of escape events with increasing $\omega_F$.}
\label{fig:prefactor}
\end{figure}

Since $\delta R(t_c)$ is a zero-mean periodic function, $\bar W$
always exceeds $W_0$. For small $F/D$, the correction to $W_0$ is
quadratic in $F/D$ (cf. \cite{Larkin}). In the opposite limit of large
$F/D$, the escape rate is changed exponentially, with $\ln[\bar
W/W_0]\approx -D^{-1}\min \delta R(t_c)$, which coincides with
Eqs.~(\ref{escape_rate}), (\ref{escape_chi}). The dependence of the
escape rate on time and the parameters of the system for a simple
metastable potential is illustrated in Fig.~\ref{fig:prefactor}.

The time dependence of the escape rate $W(t)$ and the change of its form
with varying parameters of the system, in particular with the
frequency and amplitude of the driving force, were analyzed in
Refs.~\cite{SDG,Smelyanskiy-JCP-99}. The explicit form of the
probability distribution in the vicinity of the boundary $q_b(t)$ was
obtained in these papers, too. One of the conclusions which follows
from the results is that the prefactor in the expression for the
current $j(t)$ calculated right on the boundary $q_b(t)$ has a totally
different form from that in the current well behind $q_b(t)$, which
gives the observable rate $W(t)$ (\ref{current}). This is in contrast
with what happens in the case of nondriven overdamped systems \cite{Kramers}.

\begin{figure}
\begin{center}
\epsfxsize=5.0in                
\leavevmode\epsfbox{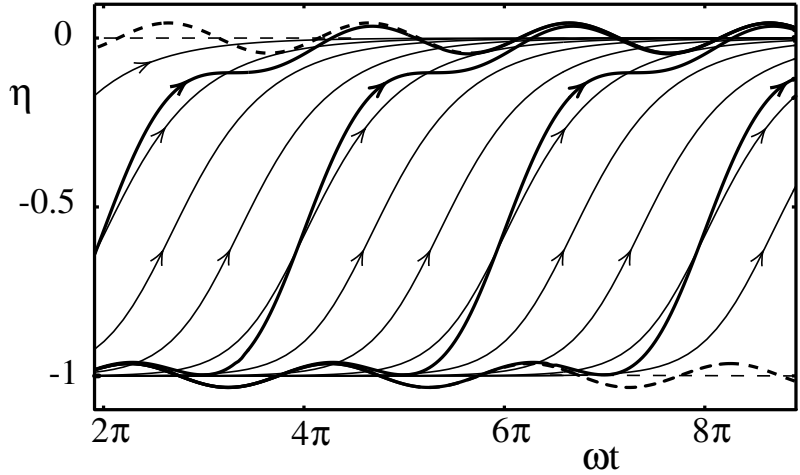}
\end{center}

\hfill

\caption{Optimal escape paths (bold solid lines) of a periodically
driven Brownian particle, $\dot \eta = \eta- \eta^3+A\cos\omega t$ $+
f(t)$, for $A=0.1$, $\omega=2$ [from Ref.~\protect\cite{Dykman:97b};
$\eta$ and $\omega$ correspond to $q$ and $\omega_F$ in the present
paper, respectively]. The paths [given by
Eqs.~(\protect\ref{boundary})-- (\protect\ref{white_noise_R})]
go from the stable to the unstable
periodic states shown by bold dashed lines (by thin dashed lines, in
the absence of driving). Thin solid lines show optimal paths in the
absence of driving $\eta^{(0)}(t-t_c)=-\{1+\exp[2(t-t_c)]\}^{-1/2}$,
with different $t_c$.  The driving lifts the degeneracy with respect
to $t_c$. The paths $\eta^{(0)}(t-t_c)$ with the ``right'' $t_c$ [as
given by (\protect\ref{escape_chi})] are the ones around which the
exact paths are oscillating.  The linear nonadiabatic theory gives the
decrement of the activation barrier to an accuracy 12\%.  }
\label{fig:Hanggi}
\end{figure}

Calculating the current at the periodic boundary $q_b(t)$ was the goal of
the recent papers by Lehmann {\it et al.} \cite{Hanggi-00}. As noted
before, the functional form of this current differs from that of the
coordinate-independent instantaneous escape rate. In their analysis,
Lehmann {\it et al.}  adopted the idea \cite{Dykman:97a,Dykman:97b},
Fig.~\ref{fig:Hanggi}, of synchronization of optimal paths by a
periodic field. The evaluation of the prefactor in \cite{Hanggi-00} is
based on an additional specific conjecture. Most of the specific
results refer to a {\it singular} potential $U(q)$ in
Eq.~(\ref{force}): it consists of two opposite-sign parabolas 
matched between their extrema. However, the nonanalyticity of this
potential should give rise to a deviation from the linear amplitude
dependence of the activation energy (\ref{escape_chi}) for
comparatively small amplitudes of the driving periodic force. The
deviation will be strong where the amplitude of forced vibrations
becomes comparable to the distance from the extrema of $U(q)$ to the
singular point where the parabolas are connected, as was indeed
observed in Ref.~\cite{Hanggi-00}. However, as we showed earlier by
solving the variational problem (\ref{escape_rate}),
(\ref{white_noise_R}) exactly \cite{Dykman:97b} (cf.
Fig.~\ref{fig:Hanggi}), for generic analytic potentials the activation
energy of escape is well described by the LS in a broad range of field
amplitudes. We demonstrate this below by analog and digital
simulations.

\section{Analog and digital simulations}

\subsection{Measuring the logarithmic susceptibility}

To test the relevance of the LS and to investigate its properties, we
have built an analog electronic model~\cite{rpp} of the system
(\ref{Langevin}) for the double-well Duffing potential 
\begin{equation}
\label{Duffing_pot}
U(q)= -{1\over 2}q^2 +{1\over 4}q^4. 
\end{equation}
We~drive it with zero-mean quasi-white Gaussian noise from a
shift-register noise generator, digitize the response $q(t)$, and
analyze it with a digital data processor. We~have also carried out a
complementary digital simulation see Ref.~\cite{riccardo97} for details
on the algorithm used and the noise generation. The analog and digital
measurements involved noise intensities in the ranges $D=0.021 - 0.04$
and $D=0.007 - 0.030$ respectively, in dimensionless units.

\begin{figure}
\epsfxsize=3.5in                
\leavevmode\epsfbox{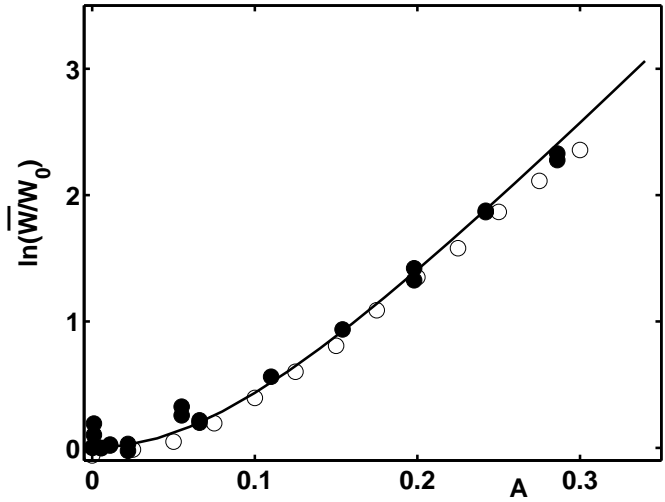}
\vspace{0.1in}
\caption{The average escape rate $\bar W$ for a sinusoidally-driven
Duffing oscillator \protect(\ref{Duffing_pot}) as a function of the
field amplitude $A$, $W_0$ is the escape rate for $A=0$.  The driving
frequency is $\omega_F=1.2$, the white-noise intensity is
$D=0.03$. Solid line: the theoretical expression
\protect(\ref{average}); filled and empty circles are the data from
analog and digital simulations, respectively, with no adjustable
parameters.}
\label{fig:data_prefactor}
\end{figure}

For escape from the state $q_a=-1$ of the white-noise driven Duffing
oscillator, Eqs.~(\ref{white-explicit}), (\ref{Duffing_pot}) give the
LS as

\begin{equation}
\label{Duffing-LS}
\tilde\chi(\omega) = \pi^{-1/2}\Gamma[(1-i\omega)/2]\Gamma[(2+i\omega)/2)],
\end{equation}
where $\Gamma(x)$ is the Gamma function. For sinusoidal driving, the
measured time-averaged escape rate is compared with the expressions
(\ref{average}), (\ref{Duffing-LS}) in
Fig.~\ref{fig:data_prefactor}. We emphasize that the data refer to a
strongly nonadiabatic driving, $\omega_F t_{\rm rel} = 0.6$ (for the
model (\ref{Duffing_pot}), $t_{\rm rel}=1/U''(q_a)=1/2$), and cover
the range from weak fields, $A \alt D$, to $A/D = 10$. The
corresponding change of $|\delta R|/D = |\tilde \chi(\omega_F)|A/D$ was $\alt
4.2$. The data and the theory are in full agreement, without any
adjustable parameters.  It is seen from the data that, for
$|\tilde\chi(\omega_F)|A/D > 1$ the dependence of $\ln \bar W$ on $A$
becomes linear, as expected. We note that a qualitatively similar
dependence of $\ln\bar W$ on the driving amplitude
can be seen in the
experimental data on driven Josephson junctions \cite{Devoret}.

\begin{figure}
\centering 
{\leavevmode\epsfxsize=3.5in\epsfysize=3.0in
\epsfbox{
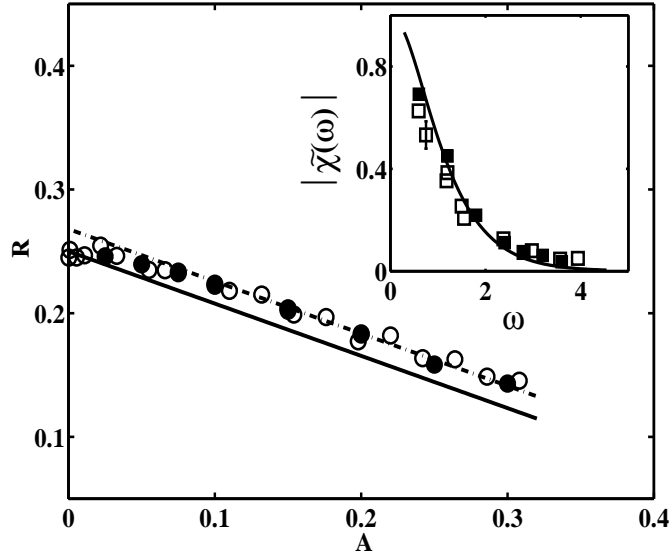}}

\hfill

\caption{The dependence of the activation energy $R$ on the amplitude
$A$ of the sinusoidal driving force with $\omega_F=1.2$ for the
Duffing oscillator as determined by electronic (open circles) and
numerical (filled circles) simulations and Eq.~(\protect\ref{average})
(solid line) \protect\cite{Luch:jpa}. The data of analog and digital simulations refer to the noise intensities $0.028 < D<0.036$
and $0.020 < D<0.028$, respectively. The inset shows the absolute value
of the LS of the system $|\tilde\chi(\omega)|$ measured (open and
filled squares for analog and numerical simulations, respectively) and
calculated from (\protect\ref{Duffing-LS}) [full curve] as a function
of frequency $\omega$.}
\label{fig:data_LS}
\end{figure}

In Fig.~\ref{fig:data_LS} we show the data on the LS for several noise
intensities. The activation energy $R$ was obtained by measuring the
slope of $\ln \bar W$ vs $1/D$. From (\ref{escape_chi}), the slope of
$R$ vs $A$ yields the absolute value of the LS. The difference between
the measured and calculated $R$ arises from the noise intensity being
not too small ($D\approx 0.020 - 0.036$ for the data points in
Fig.~\ref{fig:data_LS}), or in other words, comes from the field
dependence of the prefactor in the expression for the escape rate
$\bar W$. As seen from Fig.~\ref{fig:data_prefactor}, when the latter
is taken into account, there is full quantitative agreement between
the theory and simulations. It is also seen from
Fig.~\ref{fig:data_prefactor} that simulations with $D$ as high as $0.03$
still give the correct {\it slope} of $\delta R$ vs $A$ for large $A$,
and thus the correct $|\tilde \chi|$.

The frequency dependence of $|\tilde \chi(\omega)|$, a fundamental
characteristic of the original equilibrium system, is compared with
the theoretical prediction (\ref{Duffing-LS}) in the inset. As
expected, the LS falls off exponentially at high frequencies, whereas
the limit of $\tilde \chi(\omega)$ for $\omega\to 0$ corresponds to adiabatic
driving and can be obtained from the Kramers theory. We note that,
generally, the LS is not a monotonic function of frequency: for
underdamped systems, it displays resonant peaks \cite{Dykman:97a}.

\subsection{Switching between optimal paths}

We now turn to the investigation of a specific feature of the escape
rate that is related to the minimization over $t_c$ in (\ref{escape_chi}),
It is expected to arise for a nonsinusoidal field, and in
particular for a biharmonic one \cite{Dykman:97a}. Here, the periodic
function $\delta R(t_c)$ may have {\it two} minima per
period. However, the activation energy will always correspond to the
{\it absolute} minimum of $\delta R(t_c)$. For a certain relation
between the parameters, the values of $\delta R(t_c)$ at the two
minima are equal.  The situation is then similar to the first-order
phase transition where two minima of the free energy are equally
deep. On the opposite sides of the phase transition line the system is
in different states. In the present case, if the parameters pass
through critical values where the minima of $\delta R(t_c)$ are
equally deep, switching will occur from one minimum to the other.

For biharmonic driving, a convenient control parameter is the phase
difference $\phi_{12}$ between the field components $F_1, F_2$. In the
simulations we used the Duffing oscillator (\ref{Duffing_pot}) driven
by the field $F(t) = 0.1\cos(1.2 t) + 0.3 \cos(2.4 t +
\phi_{12})$. For such a field, the function $\delta R(t_c)$
(\ref{escape_chi}) has two minima. Their relative depths depend on
$\phi_{12}$.

\begin{figure}
\epsfxsize=6.5in                
\leavevmode\epsfbox{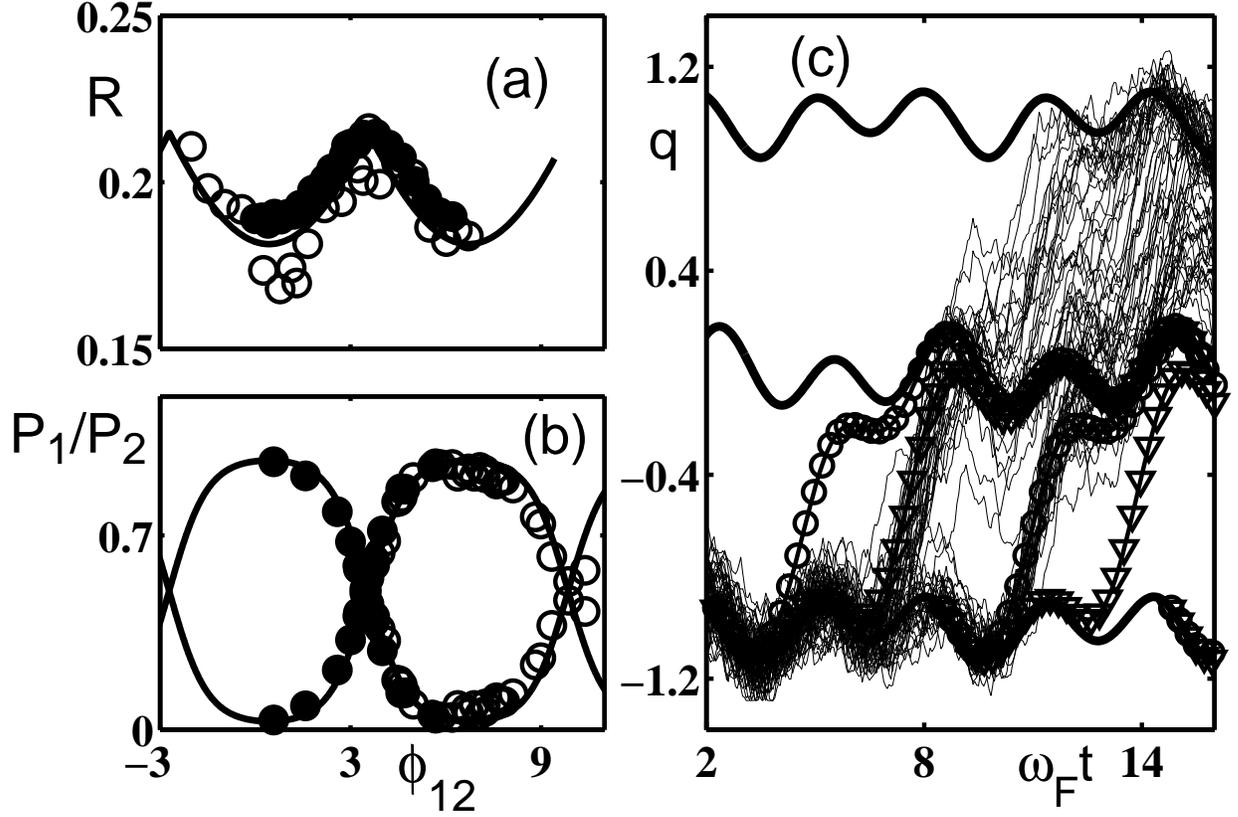}
\vspace{0.1in}
\caption{(a) The activation energy $R$ as a function of phase
difference $\phi_{12}$ with $\omega_F = 1.2$ for the Duffing
oscillator driven by the biharmonic force $F(t) = 0.1\cos(\omega_Ft) +
0.3 \cos(2\omega_Ft + \phi_{12})$.  Calculations based on
(\protect\ref{escape_chi}) (full curve) are compared with data from
electronic (open circles) and numerical (filled circles)
simulations. (b) Relative numbers of escape trajectories following
each escape path in the electronic (filled circles) and numerical
(open circles) experiments compared to the calculated relative
probabilities (full curve). (c) Measured escape trajectories for the
electronic model (thin jagged lines) with the critical phase
difference $\phi_{12}=\phi_{cr}\approx 3.57$, compared to the
calculated optimal paths (circles and triangles); solid lines are
periodic states of (\protect\ref{Langevin}), (\protect\ref{force}) in
the absence of noise. The data was obtained with the noise intensity
$D=0.028$.}
\label{fig:biharmonic}
\end{figure}

The increment of the activation energy $\delta R = \min \delta R(t_c)$
as a function of $\phi_{12}$ obtained from analog experiments and
numerical simulations is compared to theoretical predictions in
Fig.~\ref{fig:biharmonic}a. For the critical value
$\phi_{12}=\phi_{\rm cr}$, $\delta R$ has a cusp. On the opposite
sides of the cusp it is determined by different minima of $\delta
R(t_c)$. Relative numbers of escape events along the paths
corresponding to these minima are shown in
Fig.~\ref{fig:biharmonic}b. The data clearly show that the
contribution from one of the minima dominates everywhere except within
a narrow vicinity of $\phi_{\rm cr}$, where the contributions from the
both minima are of the same order of magnitude.

In Fig.~\ref{fig:biharmonic}(c) we compare observed and predicted
escape paths for $\phi_{12}=\phi_{\rm cr}$ (in the calculations,
account was taken of the field-induced corrections).  The coexistence
of the two escape paths per period is clearly seen, and agreement with
theory is excellent.

\section{Dynamical symmetry breaking in a modulated bistable optical trap}

\begin{figure}
\centering
\epsfxsize=3.3in                
\leavevmode\epsfbox{
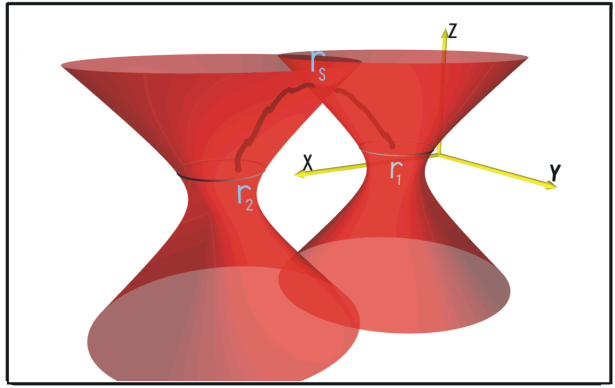}

\hfill

\caption[]{ Rendering of two
focused laser beams, the equilibrium positions of the particle (rings), and
a transitional path between the beams}
\label{fig:double_beam}
\end{figure}

A simple physical system which embodies fluctuation-induced escape is
a mesoscopic particle suspended in a liquid and confined within a
metastable potential well.  The particle moves at random within the
well until a large fluctuation propels it over an energy barrier.  An
optically transparent dielectric sphere can be readily trapped with a
strongly focused laser beam, creating an optical gradient trap,
i.e. ``optical tweezers'' \cite{Ashkin}. Techniques based on optical
tweezers have found broad applications in contactless manipulation of
objects such as atoms, colloidal particles, and biological
materials. Fluctuation-induced escape can be studied using a dual
optical trap generated by two closely spaced parallel light beams, as
illustrated in Fig.~\ref{fig:double_beam}. Such trap was implemented
initially to study the synchronization of interwell transitions by
low-frequency (adiabatic) sinusoidal forcing \cite{Simon}.

An important experiment with a particle in a double-well trap is a
measurement of the transition rate in a stationary potential. Such an
experiment can provide a rigorous test of the multidimensional Kramers
rate theory with no adjustable parameters. Quantitative measurements
require that the confining potential be adequately characterized. This
can be done by measuring directly the full
three-dimensional (3D) stationary probability distribution $\rho({\bf
r})$ of a trapped Brownian particle \cite{McCann99}.

A stable three-dimensional trap is produced by two focussed laser
beams as a result of the electric field gradient forces exerted on a
transparent dielectric spherical silica particle of diameter $2R =
0.6~\mu$m.  Displaced typically by 0.25 to 0.45~$\mu$m, the beams
create a double-well potential, with the stable positions of the
particle centered at ${\bf r}_1$ and ${\bf r}_2$.  The stability
perpendicular to the beam axis is due to the transverse beam
profile gradient; in the beam direction the potential gradient is
derived from the strong focusing of the objective lens
\cite{Ashkin}. Relatively infrequent thermally activated random
transitions between the potential wells occur through a saddle point
at ${\bf r}_s$ as depicted in Fig.~\ref{fig:double_beam}. The
experimental setup and the measurement technique have been discussed
elsewhere \cite{McCann99}.

\begin{figure}
\centering
\epsfxsize=2.5in                
\leavevmode\epsfbox{
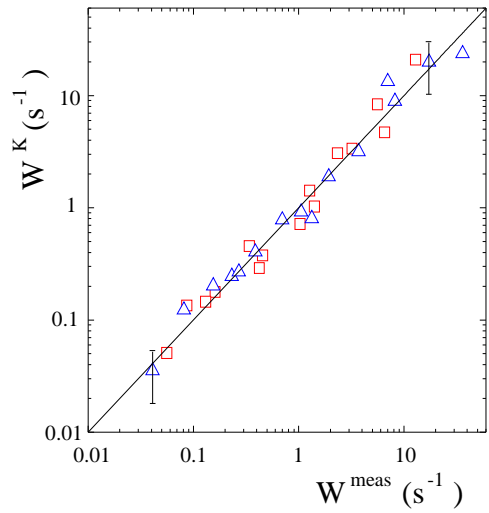}

\hfill

\caption[]{Comparison of the measured transition rates $W^{\rm meas}$ and
the rates calculated from the three-dimensional Kramers theory, $W^K$,
using the measured curvatures of the potential wells.  The squares
represent escapes from the well at ${\bf r}_1$ and the triangles
represent escapes from the well at ${\bf r}_2$ in
Fig.~\protect\ref{fig:double_beam}.  The line of slope one indicates
the result expected if the three-dimensional Kramers theory correctly
predicted the measured transition rates \protect\cite{McCann99}.}
\label{fig:kramers_rates}
\end{figure}

The full double-well confining potential $U({\bf r})$ is determined
from the measured stationary distribution $\rho({\bf r})$ as $U({\bf
r}) = -k_BT\ln \rho({\bf r})$. From the depths and curvatures of the
potential wells and the curvature of $U({\bf r})$ at the saddle point
${\bf r}_s$, it is straightforward to {\it calculate} the Kramers
escape rates. These rates can also be measured directly by placing the
particle into one of the wells and measuring the average time it takes
to switch to the other well. The potential $U({\bf r})$, and the
barrier height in particular, can be systematically varied by changing
the beam intensities. This results in an exponential change of the
escape rate, thus making it possible to compare theory and
experiment over a wide range of the escape rates. Extremely good
agreement is obtained, as seen from Fig.~\ref{fig:kramers_rates}.

The double-beam trap can also be used to investigate the effect of
ac-modulation on transition rates. An interesting application of this
effect is to {\it direct} the diffusion of a particle in a spatially
periodic potential \cite{Magnasco}.  It follows from the results of
Sec.~2 that, for a generic periodic potential, the ac-induced change of
the activation barrier differs depending on the direction (right
or left, for example) in which the particle moves in escape. This
makes the probabilities of transitions to the right and to the left
exponentially different and results in diffusion in the direction of
more frequent transitions.

An effect closely related to directed diffusion, but more amenable to
testing using optical trapping, is ac-field induced {\it localization}
in one of the wells of a symmetric double-well potential. We expect
both these effects to occur if the applied field breaks the
spatio-temporal symmetry of the system \cite{Dykman:97a},
\cite{Ajdari}. The ratio of the stationary populations $\bar w_1, \bar
w_2$ of the wells is determined by the ratio of the period-averaged
rates $\bar W_{ij}$ of the interwell $i \to j$ transitions,
\begin{equation} 
\bar w_1/\bar w_2 = \bar W_{21}/\bar W_{12} \propto \exp(\left[\delta
R_1-\delta R_2\right]/k_BT),
\label{population_ratio}
\end{equation}
\noindent where $\delta R_{1,2}$ are field-induced corrections to the
activation energies of escape from wells 1,2 (\ref{escape_chi}).

\begin{figure}
\centering
\epsfxsize=3.0in                
\leavevmode\epsfbox{
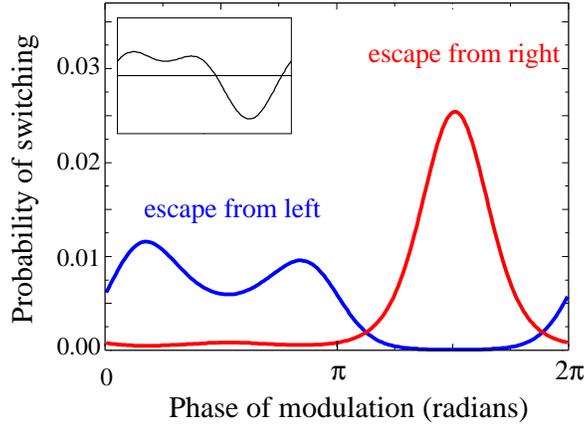}

\hfill

\caption{The least-squares fits to the experimentally determined
instantaneous time-dependent switching probabilities for a particle in
the adiabatically modulated double-beam trap, over a cycle $\omega_F
t$ of the modulating waveform.  The phase angle between the first and
second harmonics is $\phi_{12}=\pi/2$. When the phase angle was
incremented by $\pi$, the escape rates from the left and right wells
interchanged, within experimental error. Inset shows the instantaneous
difference between the heights of the potential barriers in the two
wells.}
\label{fig:adiabatic}
\end{figure}

The experiment was conducted \cite{MDG_to_be} for equal static barrier
heights in the two wells $\Delta U_1=\Delta U_2 \equiv \Delta U$, with
$\Delta U_0$ set at $ \approx 7.5 k_BT$. The intensity of a laser beam
was then modulated by an electro-optic device, giving rise to
modulation of $\Delta U/k_BT$ with an amplitude $\approx 2.5$. The
modulation frequency $\omega_F/2\pi$ was varied between 1 and 100 Hz,
which covers the range from adiabatically slow to nonadiabatic
modulation [the relaxation time is $t_{\rm rel}\sim 10^{-2}$~s].  This
may be compared to the mean unmodulated transition rate $W_0 \sim 0.1\,
{\rm s}^{-1}$. Over this range, field-induced re-population was
observed between the wells for a nonsinusoidal modulation waveform, so
that $\bar w_1\neq \bar w_2$,

The results on the instantaneous escape rates $W_{12}(t)$ and
$W_{21}(t)$ for an adiabatic modulation ($\omega_F/2\pi = 1$~Hz) are
shown in Fig.~\ref{fig:adiabatic}. The barrier heights in the two
wells were modulated in counter-phase. The form of the modulation was
$\delta \Delta U(t) = {\rm const}\times [\sin(\omega_F t) +
(1/2)\sin(2\omega_F t + \phi_{12})]$. For this waveform, there is only
one optimal escape path per period, for each well, and no switching
between the paths occurred with varying $\phi_{12}$.

As shown in the inset to Fig.~\ref{fig:adiabatic}, the difference
between the barrier heights in the two wells varies asymmetrically
over the cycle. It depends on $\phi_{12}$ and can be inverted if the
phase angle is shifted by $\pi$. In other words, the modulated
potential is not invariant under $t\to t+\pi/\omega_F, x\to -x, y\to
-y$ (with $x,y$ measured from the symmetry planes parallel to the beam
axes, see Fig.~\ref{fig:double_beam}). It is this breaking of the
spatio-temporal symmetry that leads to the escape rate from one of the
wells being on average much bigger than from the other, as seen from
Fig.~\ref{fig:adiabatic}. In turn, this leads to a higher population
in one of the wells. Not only has the effect been observed for slow
modulation, as evidenced by Fig.~\ref{fig:adiabatic}, but a population
difference of $20\%$ has been observed deeply in the nonadiabatic
regime, with $\omega_F/2\pi = 20$~Hz, for the modulation amplitude
used.  This is sufficient to create significant directional diffusion,
and demonstrates the onset of {\em dynamical} symmetry breaking. The
dependence of $\bar w_1/\bar w_2$ on the phase shift $\phi_{12}$ is in
agreement with the theory of Sec.~III \cite{MDG_to_be}.

\section{Conclusions and Open Questions}

We have shown theoretically, by analog and digital simulations, and by
optical trapping experiments that fluctuations in driven systems, and
in particular escape from a metastable state, can be effectively
controlled by an external field.  The field gives rise to a change of
the activation energy of escape, which can be much bigger than the
characteristic noise intensity (temperature) even for comparatively
weak fields. Over a broad range of field amplitudes, this change is
{\it linear} in the field even where the driving frequency exceeds the
reciprocal relaxation time of the system and substantially exceeds the
escape rate. The effect is described by a physically observable
quantity, the logarithmic susceptibility. The LS relates the
probability of large fluctuations in the presence of an external field
to the dynamics in the absence of driving. It displays a specific
frequency dispersion, which makes it possible to control, selectively,
the escape rate of a targeted system. The LS can be calculated, for a
given model of the system, and can  be measured experimentally.

An important application of the escape rate control is directed
diffusion in a spatially periodic potential. As we have demonstrated,
such control can be performed even for a symmetric potential, in which
case not only the rate, but also the preferred direction of diffusion
can be conveniently changed by changing the driving field parameters.

The high efficiency of the escape rate modulation is largely due to
the {\it synchronization} of optimal escape paths by the driving field.
We have predicted and observed this synchronization. We have also
observed the related effect of switching between different branches of
the activation energy as a function of the field parameters -- a
generic phase-transition type effect related to coexistence of
different escape paths in systems away from thermal equilibrium.

The results of this research are relevant to biological systems, since
activated escape lies at the root of many biological processes at the
molecular level, and modulation of the escape rate is often the way
nature excercises control. However, detailed understanding of how this
control is performed is missing in most cases. This is a fundamentally
important and most challenging open scientific problem. 

Another important open problem of broad interest is whether large
fluctuations can be used to learn about the dynamics of a fluctuating
system away from stable or metastable states. The underlying idea here
is that, in large fluctuations, the system explores remote areas of
the space of its dynamical variables. An example where fluctuations
have been used to find the global potential in which a system is
moving was discussed in Sec.~V. The problem becomes more complicated
if the system dynamics is unknown. However, given that the system is
most likely to move along a certain path during a large fluctuation,
the observation of such paths can enable one to infer a dynamical
model of the system. The process can then be iterated, as indicated by
recent results \cite{Vadim_infer}.


The results described here are of interest also from the viewpoint of
practical applications.  An important example is the separation of
colloidal particles and macromolecules. Our experiments show that
selectively directed diffusion is a promising new approach to this
problem. Another example is control of crystal growth using an ac
field. The relevant nucleation rate will be changed by the field in a
way similar to the escape rate. Therefore it should be possible to
strongly modulate it both in time and space.

The work was supported by the NSF through grants no.~PHY-0071059 and
DMR-9971537, and by the EPSRC (UK) under grants Nos.\ GR/L99562 and
GR/R03631.

\newpage

\newpage

\newpage

\end{document}